\def\cm3{cm$^{-3}$}
\def\kms{km~s$^{-1}$}

\def\lsun{L$_{\odot}$}

\def\msun{M$_{\odot}$}

\def\beq{\begin{equation}}
\def\eeq{\end{equation}}

\input{aipcheck}

\documentclass[
    ,final            
  ]
  {aipproc}

\layoutstyle{6x9}

\begin{document}

\title{Time-dependence Effects in Photospheric-Phase Type II Supernova Spectra}


\author{Luc Dessart}{
  address={Steward Observatory, The University of Arizona, Tucson, AZ \ 85721, USA}
}

\author{D. John Hillier}{
  address={Department of Physics and Astronomy, University of Pittsburgh, USA}
}

\begin{abstract}
We have incorporated time-dependent terms into the statistical and
radiative equilibrium calculations of the non-LTE line-blanketed radiative
transfer code CMFGEN. To illustrate the significant improvements in
spectral fitting achieved for photospheric phase Type II SN, and to
document the effects associated with time dependence, we model the
outer 6.1\,\msun\, of ejecta of a BSG/RSG progenitor star. Hopping by
3-day increments, we compute the UV to near-IR spectral evolution for
both continuum and lines, from the fully ionized conditions at one week
to the partially recombined conditions at 6 weeks after the explosion.
We confirm the importance of allowing for time-dependence in the modeling of 
Type-II SN, as recently discussed by Utrobin \& Chugai for SN1987A.
However unlike Utrobin \& Chugai, who treated the radiation field in a
core-halo approximation and assumed the Sobolev approximation for line
formation, we allow for the full interaction between the radiation field
and level populations, and study the effects on the full spectrum.  At the
hydrogen-recombination epoch, H{\sc i} lines and NaD are considerably
stronger and broader than in equivalent steady-state models, while
Ca{\sc ii} is weakened.  Former successes of steady-state CMFGEN models
are unaffected, while former discrepancies are cured. Time dependence
affects all lines, while the continuum, from the UV to the optical,
changes only moderately.  We identify two key effects: First, time
dependence together with the energy gain through changes in ionization and
excitation lead to an over-ionization in the vicinity of the photosphere,
dramatically affecting line optical depths and profiles. Second, the
ionization is frozen-in at large radii/velocities. This stems solely
from the time-scale contrast between recombination and expansion
and will occur, modulo non-thermal excitation effects, in all SN types.
The importance of this effect on spectral analyses, across SN types and 
epochs, remains to be determined.
\end{abstract}

\maketitle

\section{Introduction}

Because of radiative cooling and the fast expansion of the exploding mantle of the progenitor massive
star, photospheric-phase Type II supernova (SN) spectra evolve rapidly.
At shock breakout
the spectral energy distribution (SED) peaks in the far-UV and X-rays. Subsequently the SED
centers in the UV, then in the optical, and later in the infrared until the object disappears altogether into
oblivion after a few years. From a radiative transfer perspective, the cooling induces
a recombination to lower ionization stages that impose a strong blanketing effect on the energy
distribution. A few days after explosion, the UV/optical/IR spectrum shows a UV-dominant SED,
with Balmer/Paschen lines of Hydrogen, He{\sc i}\,5875\AA, He{\sc i}\,10830\AA, N{\sc ii}\,4600\AA,
N{\sc ii}\,5400\AA, and a few isolated resonance lines such as Mg{\sc ii}\,2802\AA\ and Al{\sc iii}\,1859\AA.
About a week later, hydrogen recombination kicks in, tied to a sudden enhancement in line-blanketing
due to the switch from Fe{\sc iii} to Fe{\sc ii}. \citet{DH2006} successfully modeled
these epochs in the evolution of SN1999em, using steady-state non-LTE CMFGEN models,
solar metallicity, and a power-law density decline with an exponent of ten.
In particular, He{\sc i} lines at early times were fitted without invoking
any additional non-thermal excitation due to $^{56}$Ni, as required by \citet{Baron_etal2003}
in their analysis of SN1993W.
In \citet{DH2006}, the spectroscopic analysis deliberately focused on the
first 40 days since at such an epoch, we encountered growing difficulties in reproducing
H$\alpha$, H$\beta$, and the near-IR Paschen lines \citep{Hamuy_etal2001}.
While we could reproduce the overall SED very well, line profiles were sizably narrower
than observed at late times, suggesting that line formation regions predicted by CMFGEN
were confined to velocities/radii that were too small. CMFGEN failed to reproduce observations,
despite all attempts, whenever hydrogen started recombining at and above the photosphere.

   The H$\alpha$ problem has not been clearly emphasized in the literature, where one can see a great disparity in
model atmosphere assumptions and agreement between theoretical predictions and observations,
but no direct links to physical/numerical issues.
In the eighties and early nineties, the recognized importance of non-LTE effects confronted the strong
limitations of computer technology, so that only a few species were treated in non-LTE
(i.e., usually hydrogen and helium), while the metals responsible for line blanketing were treated in LTE.
\citet{EastmanKirshner1989} followed this approach to model the
first ten days of SN1987A, and thus did not encounter obvious difficulties with hydrogen lines.
\citet{Schmutz_etal1990}, using an approximate non-LTE technique,
had, on the contrary, great
difficulty reproducing any of the Balmer lines, suggesting clumping as the culprit.
\citet{Hoeflich1988} reproduced the SN1987A spectral evolution and the hydrogen lines,
over many months, using a large ``turbulent'' velocity.
In the more sophisticated non-LTE CMFGEN \citep{HillierMiller1998, DH2005a}
models presented in \citet[DH06]{DH2006}, we found that decreasing the
turbulent velocity weakens line-blanketing
effects and increased, although only modestly, the strength of hydrogen lines.
However, beyond 40 days after
explosion, this tuning had no longer any important influence on the hydrogen lines.
The non-LTE model atmosphere code PHOENIX reproduces the Balmer lines well
\citep{Baron_etal2003, Mitchell_etal2001}, but this may stem from their adoption of
non-thermal ionization/excitation due to $^{56}$Ni at the photosphere just a few days after
explosion, in mass shells moving at $\ge$10000\,\kms in SN1987A \citep{Mitchell_etal2001}
or $\sim$9000\,\kms in SN1993W \citep{Baron_etal2003}.
Hydrodynamical simulations of core-collapse SN predict that $^{56}$Ni
has velocities of at most $\sim$4000\,\kms, the nickel fingers being strongly
decelerated at the H/He interface \citep{Kifonidis_etal2000, Kifonidis_etal2003}.
The magnitude of this disagreement extends far beyond the uncertainties of explosion
models and suggests a genuine incompatibility.

  Interestingly, H{\sc i} lines remain strong for months in all Type II SN, e.g.
the ``peculiar'' SN1987A, the ``plateau'' SN1999em, and the ``low-luminosity'' SN1999br.
CMFGEN models computed for these objects were unable to reproduce Balmer lines
after $\sim$4 days in SN1987A, $\sim$40 days in SN1999em, and $\sim$20 days in SN1999br, all coincident
with hydrogen recombination in the outflow.
These three SN boast very different envelope structures, ejecta velocities,
and light curves. SN1999br even synthesized an order of magnitude less $^{56}$Ni than average
\citep{Pastorello_etal2004}.
The only common property between these objects, which is connected to the H$\alpha$ problem,
is the recombination of the ejecta to a lower ionization state at the corresponding epoch.

Recently, \citet[UC05]{UtrobinChugai2005} proposed that the effect of time-dependence, and
the energy associated with changes in ionization/excitation, lead to a strong
H$\alpha$ line profile in SN1987A during the recombination epoch.
They also found that barium lines were affected, and that, with their more
consistent approach, Ba{\sc ii}\,6142\AA\ could be fitted using the LMC metallicity value.
The steady-state models of \citet{Mazzali_etal1992} supported instead an abundance enhancement
of five.
Time dependence has been invoked in the past by \citet{FranssonKozma1993} to explain the
late-time light curve of SN1987A, and the theoretical study of \citet{PintoEastman2000a, PintoEastman2000b}
showed that time dependence {\it in the radiation field} had a critical impact on the 
radiative transfer in Type Ias. \citet{PintoEastman2000a, PintoEastman2000b}, however, treat the material in 
LTE, in other words, they neglect any non-LTE and time-dependent terms in the rate equations
(i.e. they do not solve the rate equations),
focusing instead on the time-dependent {\it diffusion} of photons through an optically
thick Type Ia SN ejecta. Similarly, \citet{Kasen_etal2006} neglect such explicit time dependence in 
the rate equations, computing the ionization and excitation state of the medium in LTE.
Thus, although there is at present growing interest in accounting for time dependence in the radiation field,
the often-used expedient of LTE, to maintain CPU costs low, has forced the neglect of both non-LTE and
time-dependence in the level populations. 
One exception to these time-dependent LTE approaches is the work of \citet{Hoeflich2003}, who 
treats time-dependence in the rate equations but, to our knowledge, 
has not discussed the associated effects on Type II SN spectra.
In the present and forthcoming studies, we investigate thoroughly the effects of time-dependence 
in the rate equations and their impact on inferred ejecta properties. Time dependence in the 
radiation field is accounted for by ensuring that the emergent light from the model matches the
observed flux from the Type II SN, using SN1999em as our reference SN \citep{DH2006}.

   Here, we report the salient features of a {\it time-dependent} non-LTE
CMFGEN simulation covering the evolution of a Type II SN from one to six weeks
after core collapse.
We confirm the results of UC05 that time-dependence
induces an over-ionization of the recombining ejecta and that it solves the
H$\alpha$ problem. 
However unlike UC05, we self-consistently solve the radiation
transfer equation --- the coupling between the level populations and the radiation field
is calculated and fully allowed for. Our study of the full spectrum 
predicts that all lines are significantly affected.
The ionization of the ejecta and its evolution are so strongly modified that we
predict even He{\sc i}\,10830\AA\ many weeks after explosion. This feature has been observed
\citep{Hamuy_etal2001, Spyromilio_etal1991}, but never understood, and epitomizes the
difficulty of line identification, abundance determinations, and chemical stratification
in SN ejecta. In the next section, we present our baseline model, before
moving onto the presentation and discussion of our results.

\section{Model}
\label{sect_model}

  CMFGEN \citep{HillierMiller1998, DH2005a} solves the radiative transfer equation
in the comoving frame, subject to the constraints of radiative and statistical equilibrium.
It assumes spherical symmetry.
No explicit time-dependence in the radiation field is considered, i.e. light propagates
instantaneously and thus is not subject to travel-time delays \citep{EnsmanBurrows1992},
and we still ignore relativistic effects. Both affect the
energy transport (luminosity) at depth \citep{PintoEastman2000a, PintoEastman2000b}.
However, unlike the version of CMFGEN used in
\citep{DH2005a, DH2005b,DH2006}, we now include explicitly time dependence in the
rate equations that are solved for level populations, as well as in the energy equation
by accounting for changes in ionization and excitation. 
Our treatment, which follows directly from \citet{MihalasMihalas1984}, 
will be presented in detail in a follow-up paper (Hillier \& Dessart 2006, in prep.). 

  We adopt a different procedure to model SN spectra from that employed previously.
Rather than modeling, at each time, the region in the vicinity of the photosphere (i.e., from
$\tau_{\rm Rosseland}$ of 30), we model the time evolution of a larger
portion of the ejecta. As a starting point for our baseline model, we mapped onto the
CMFGEN grid an envelope with the following properties:
$R_0 = 2.524 \times 10^{14}$\,cm ($R_0$ is the base radius), $v_0 = 3470$\,\kms
($v_0$ is the base velocity), $R_0^3 \rho_0 = 6.781 \times 10^{33}$\,g
($\rho_0$ is the base density), $\rho(r) = \rho_0 \left( R_0/r \right)^{10}$ ($r$ is the radius),
and $L_0 = 3.6 \times 10^{11}$\lsun ($L_0$ is the base luminosity).
The maximum radius is 16 times the value of the base radius $R_0$.
These initial conditions correspond to a base Rosseland optical depth of 4000, a total
mass of 6.1\,\msun, and, assuming homology, a time after explosion of 8.4 days.
We adopt a fixed composition, i.e. chemical homogeneity, with values consistent with
the red-sugergiant progenitor of SN1999em \citep{DH2006, Smartt_etal2002}.
Adopted abundances, given by number, are: H/He = 5, C/He = 0.00017, N/He = 0.0068, O/He = 0.0001,
and all metal species are at the solar value.
Compared to the model atom used in DH06, we omit
C{\sc i}, neon, aluminium, sulfur, chromium, manganese,  and cobalt.

   An initial steady-state model is then computed with the above parameters and it gives a spectrum that
fits the observations of SN1999em at 7 days after explosion \citep[Fig.~1]{DH2005a},
with a {\it fully-ionized} ejecta, little blanketing in the UV, and a dominance of hydrogen
Balmer lines and He{\sc i}\,5875\AA\ in the optical range.
We then hop in time by steps of $\Delta t = 3$\,days, i.e., all mass shells $m$ in the initial model
are evolved from their position $r(m)$ to a new radius $r'(m)$, with $r'(m) = r(m) + v_0(m) \Delta t$,
a new density $\rho' = \rho (r/r')^3$, but the same velocity.
In practice, for each new model, CMFGEN adapts its 80-point grid to cover
each optical-depth decade with at least five points.

   The luminosity, set at the base, is adjusted at each epoch so that the
{\it emergent} luminosity matches the inferred values for SN1999em (DH06), i.e. from
$\sim$10$^9$\,\lsun\, one week to $\sim 10^8$\,\lsun\ 50 days after explosion.
Despite the fixed composition at all times and the neglect of C{\sc i}, both of which affecting the
late time appearance of Type II spectra (DH06) we support our thesis by showing
in the next section a {\it representative} match between the spectrum of our baseline model and the
optical observations of SN1999em. Detailed analyzes will be presented in a future paper.

  \section{Results and Discussion}
\label{sect_results}

\begin{figure}
  \includegraphics[height=.23\textheight]{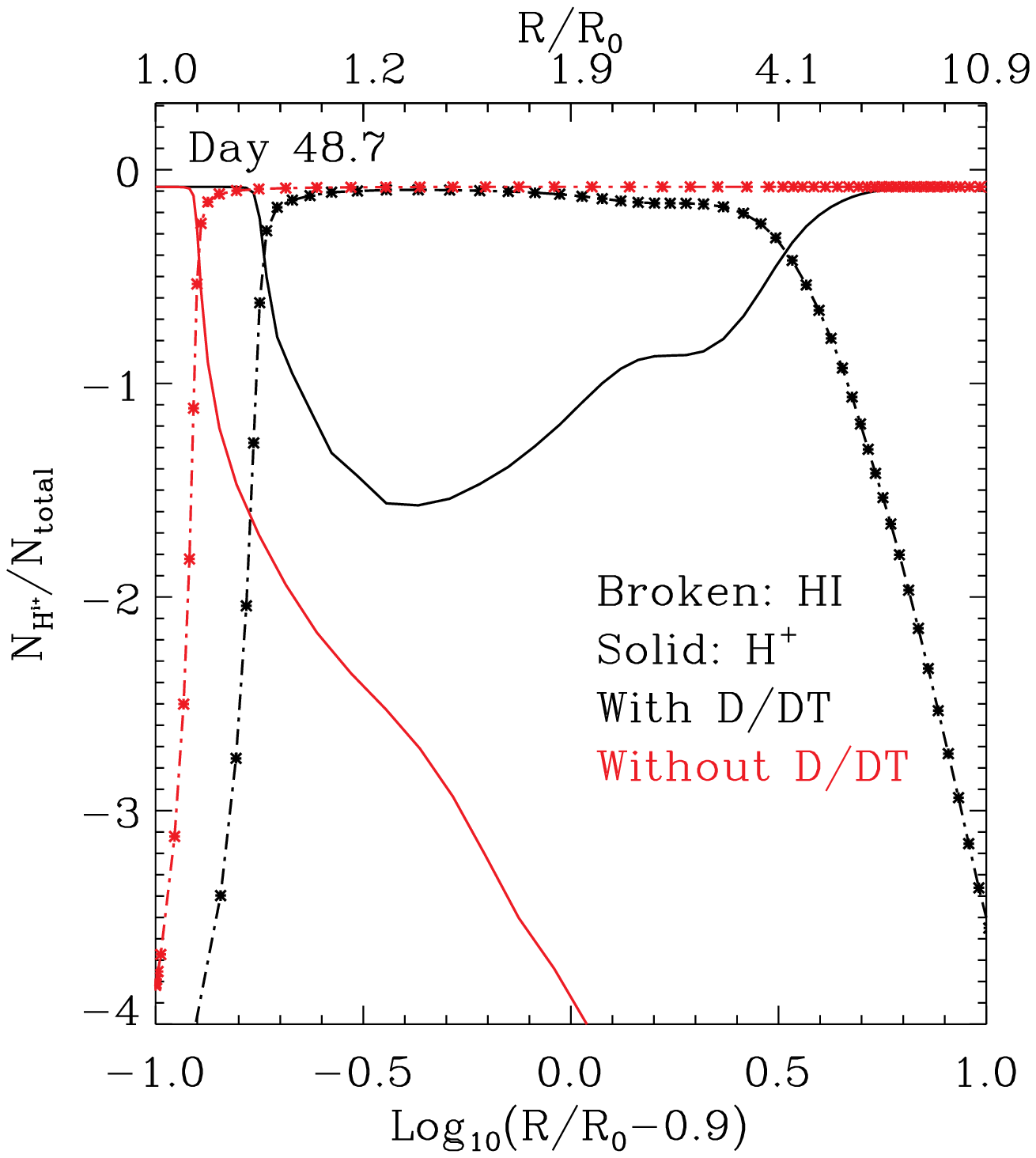}
  \includegraphics[height=.23\textheight]{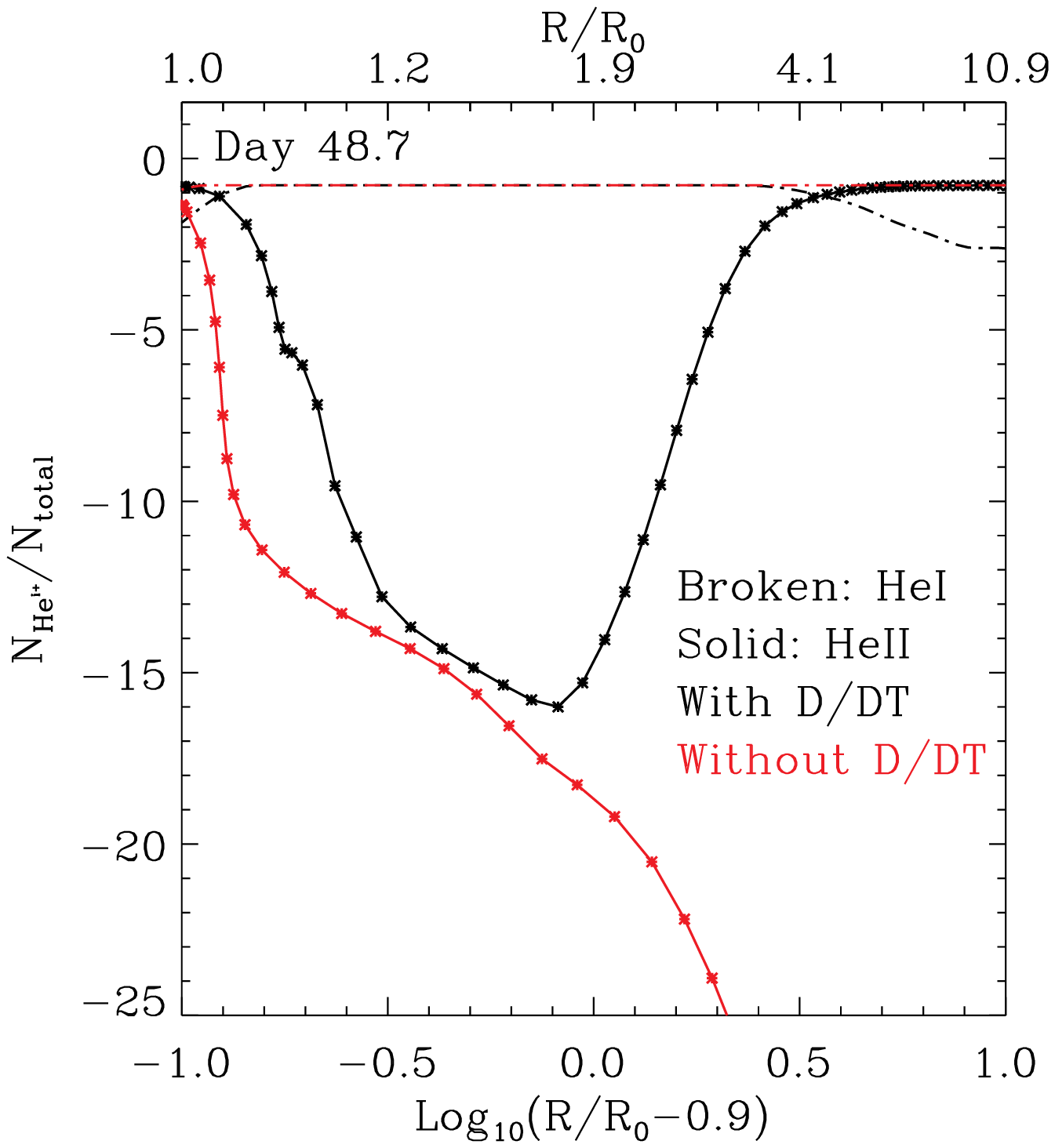}
  \includegraphics[height=.23\textheight]{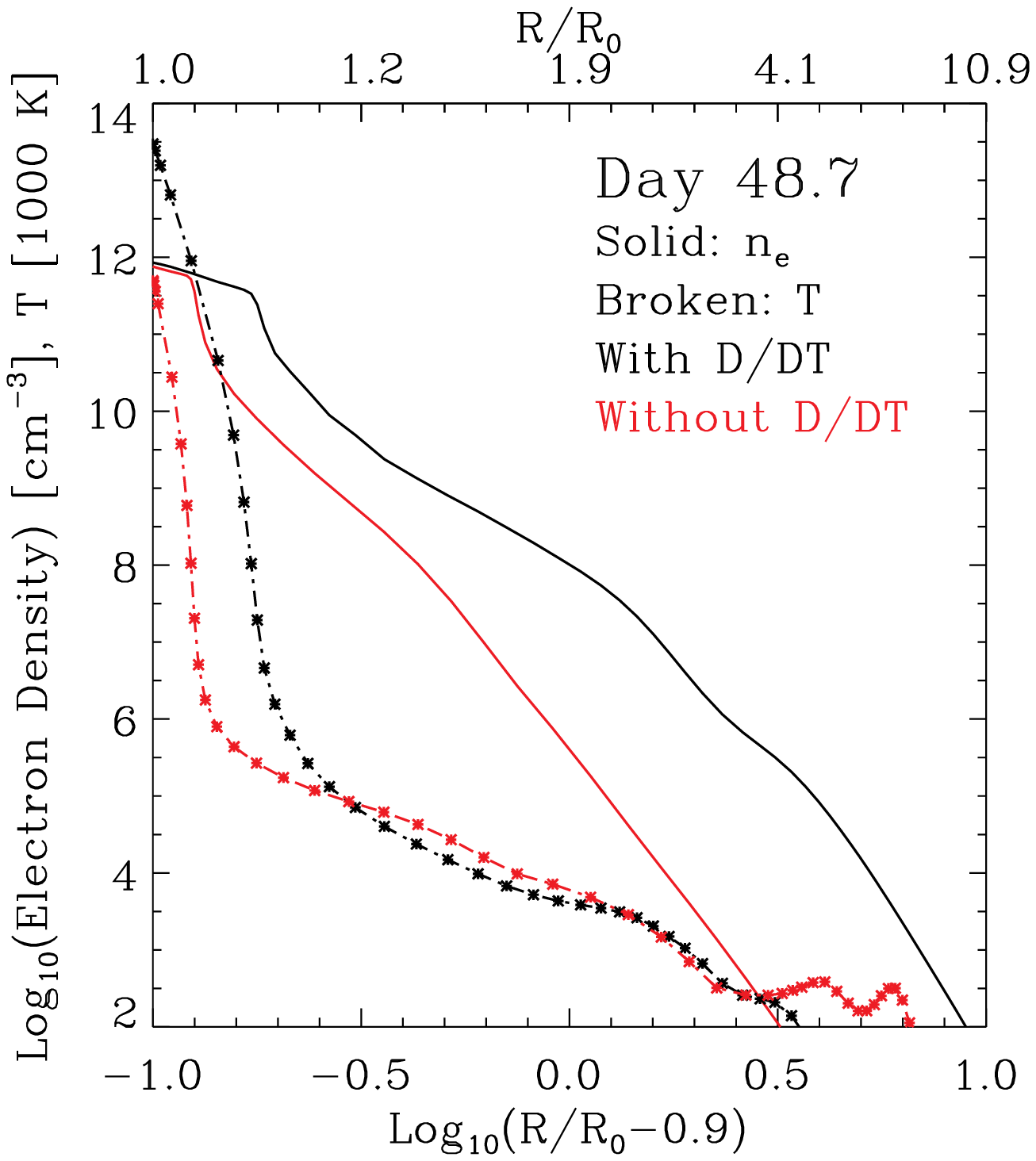}
\caption{
{\it Left}: Radial variation of the ionization fraction of Hydrogen
(H{\sc ii}: solid line; H{\sc i}: Broken line) for the models with
time-dependence (black) and without (red), and at 48.7 days after explosion,
but otherwise having {\it identical} parameters.
Note the overionization of hydrogen at large radii, not present
in its neutral state despite the very late time, akin to a frozen-in ionization.
Symbols give the positions of the CMFGEN adaptive-grid points, for each model but for only
one curve.
Note the presence of the well-resolved recombination front near the base,
deeper in for the steady-state CMFGEN model.
{\it Middle}: Same as left, but for the ionization fraction of helium
(He{\sc ii}: solid line; He{\sc i}: Broken line).
{\it Right}: Same as left, but for electron density and temperature.
}
\label{fig_pop}
\end{figure}

 To highlight the salient features of time-dependent CMFGEN models, we select the last baseline
model computed in the sequence, corresponding to 48.7\,days after explosion, and compare its
properties to the same model, but computed by assuming steady-state.
Fig.~\ref{fig_pop} shows the radial variation of
the ionization fraction of hydrogen and helium for the time-dependent
and steady-state models (left and middle panels).
Time dependence systematically reduces the effective recombination rates, with the
greatest effect occurring away from the photosphere where the densities are lowest and
the velocities highest.  In those regions and at 48.7\,days, the ejecta preserves the high ionization
it had at the start of the time sequence, at $\sim$8\,days after explosion, exhibiting
the prevalence of ionized hydrogen and helium.
Many other species are also affected, often showing comparable ionization profiles
(i.e., a reduction of their degree of ionization near the photosphere
and a growth of their degree of ionization at larger distances).
Further, because of the enhanced H ionization, the electron density does not fall as steeply
in the time-dependent CMFGEN model (Fig.~\ref{fig_pop}).
The electron temperature is strongly affected
in the optically-thick layers, which remain hotter due to the reduced radiative cooling
efficiency in the time-dependent case.

Thus, the effect of time dependence modifies the optical-depth of both continua and lines.
The former occurs since more free electrons are available to scatter continuum photons.
Further, certain ions are ``over-abundant'' by several orders of magnitude compared to the steady-state model.
In the top panel of Fig.~\ref{fig_spec_DDT_noDDT}, although not tailored specifically to match an
observation, the reddened ($E(B-V)=0.1$) spectrum of our baseline model (black) is overlaid on
the optical spectrum of SN1999em observed on the 5th of Dec 1999 \citep[blue]{Leonard_etal2002}, i.e.
$\sim$45 days after explosion (DH06). Note the strong H$\alpha$ line in this time-dependent
CMFGEN synthetic spectrum, in aggreement with the observation.
In the lower panels, we present synthetic UV, optical, and near-IR spectra for the time-dependent
and steady-state (red) CMFGEN models, normalized at 12000\AA.
To highlight the effect on individual species, we also compare
synthetic spectra computed by allowing all continuum processes, but accounting only for
bound-bound transitions of Fe{\sc ii} and H{\sc i}.
Note the large effect on Balmer and Paschen lines, which appear stronger in both absorption and
emission and are broader.
Fe{\sc ii} lines are stronger in the steady state model ---
stronger in absorption in the UV and stronger in emission in the optical.
Similarly, in the lower panel, we zoom in on specific synthetic lines, accounting only
for bound-bound transitions of Na{\sc i}, Ca{\sc ii}, or He{\sc i}.

In the time-dependent case,
Na{\sc i}\,5895\AA\ strengthens and broadens. This occurs because
Na$^+$ is the dominant ionization state in both models, but the increased
electron density in the time-dependent model leads to an increase in the
neutral Na fraction. Conversely, Ca{\sc ii}\,8500\AA\ weakens, primarily because
the relative fractions of Ca$^+$ and Ca$^{++}$ change significantly.
He{\sc i}\,1.083\,$\mu$m is present only in the time-dependent case,
as a broad flat-topped profile (overlapping with P$\gamma$ in the full spectrum; top panel
of Fig.~\ref{fig_spec_DDT_noDDT}),
with a blue absorption maximum at 1.038\,$\mu$m (equivalent to $-$13000\,\kms).

The implications of time dependence are numerous and far reaching.
We find that the effects of time dependence on level populations and on the energy budget allow
the best of two worlds. Previous successes of CMFGEN at fitting the overall spectral energy distribution
of photospheric phase Type II SN are left unchanged. The outstanding problem of the H$\alpha$ line strength
and the overall line profile widths are however solved. We, thus, confirm the importance of time-dependence
effects, as suggested by UC05, even at early times.
This provides a more physical and less dramatic explanation of the apparent ionization/excitation
seen in photospheric-phase Type II SN spectra than invoking non-thermal ionization/excitation
by $^{56}$Ni/$^{56}$Co in ejecta regions moving sometimes at $\sim$10000\,\kms.

\begin{figure}
  \includegraphics[width=.65\textwidth]{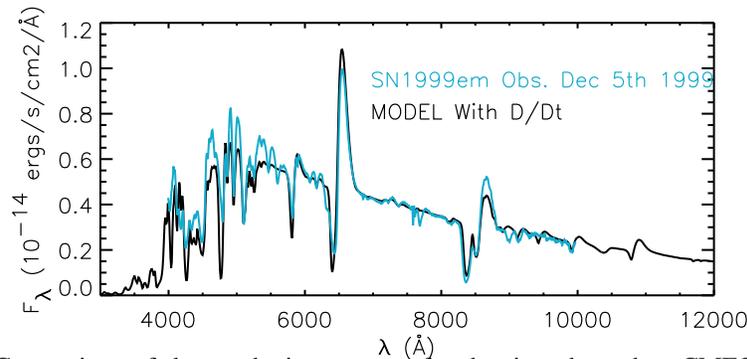}
  \caption{Comparison of the synthetic spectrum for the time-dependent CMFGEN model (black curve)
at 48.7 days after explosion  with Keck observations of SN1999em \citep{Leonard_etal2002}
on Dec. 5th 1999 (turquoise curve), at $\sim$45 days after explosion (DH06). The model
has been reddened with $E(B-V)=$0.1.}
\end{figure}
\begin{figure}
  \includegraphics[width=.65\textwidth]{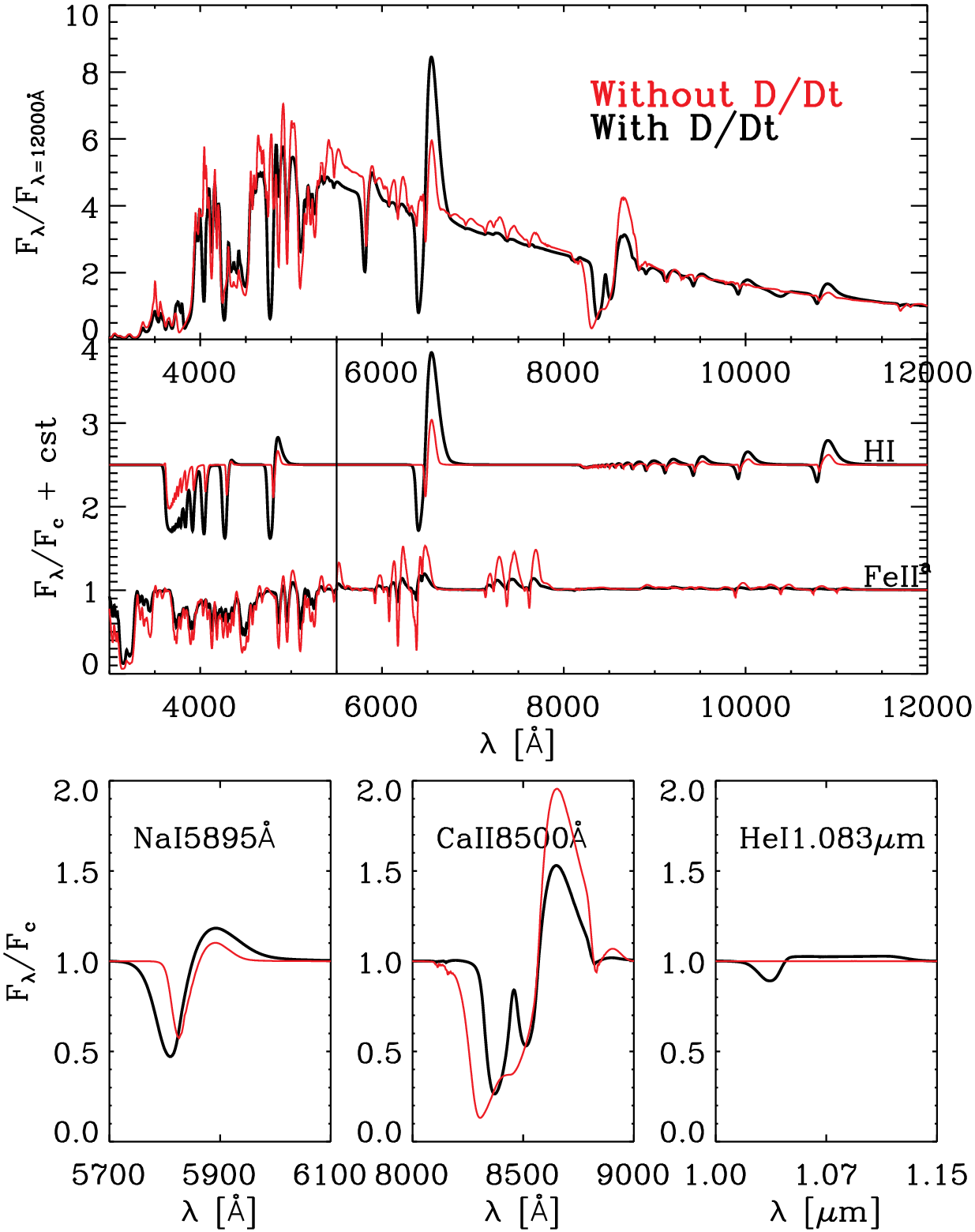}
\caption{
{\it Top}: Comparison of the synthetic spectra for the time-dependent CMFGEN model (black curve)
and the equivalent steady-state CMFGEN model (red curve), both at 48.7 days after explosion.
Accounting for time dependence leads to a comparatively higher ionization in
the ejecta, sustained line optical depth at large radii/velocities, epitomized by H$\alpha$.
{\it Middle}: Rectified spectra (shifted vertically) obtained by accounting solely for
bound-bound transitions of Fe{\sc ii} (lower curves; line spectrum magnified by a factor of four
beyond 5500\AA\ for better rendering) and H{\sc i} (upper curves).
{\it Bottom}: Same as above, but this time zooming-in on specific
spectral regions, i.e., Na{\sc i}\,5895\AA\ (left), Ca{\sc ii}\,8500\AA\ (middle),
and He{\sc i}\,1.083$\mu$m (right).
}
\label{fig_spec_DDT_noDDT}
\end{figure}

The change in ionization is large.
Time dependence affects line profiles in a way that can mimic changes in chemical
abundance in steady-state
models, with dire consequences on the quantitative assessment of the chemical composition
of SN outflows. Even more extreme is the case of He{\sc i}, absent
in our steady state models after $\sim$2 weeks, but present in equivalent time-dependent models
even after 6 weeks, without any enhancement in abundance from the initial red-supergiant
composition. An unidentified feature at 1.03$\mu$m was observed by \citet{Hamuy_etal2001} in SN1999em
at 4 weeks after explosion but not reproduced by DH06 with steady-state
CMFGEN models; with time dependence effects, it can now be
identified as He{\sc i}\,1.083\,$\mu$m.
It also seems to be present in SN1987A a few weeks after the explosion \citep{Spyromilio_etal1991}.
It has been inferred in Type Ib/c SN \citep{Hamuy_etal2002}, produced by non-thermal
excitation \citep{Lucy1991}, and even suggested in Type Ia SN \citep{MazzaliLucy1998} 
, although Mg{\sc ii} is expected to contribute, perhaps dominantly,  
to the feature \citep{Marion_etal2003}.
In the future, we will investigate to what extent time dependence can be
distinguished from non-thermal excitation, and if, together with the accurate treatment of
non-LTE effects, they can help resolve the issue of the presence of helium in SN spectra.

Our investigation suggests that to model the emergent light from a SN
at a given time, one needs to have the knowledge of the physical conditions in the SN
envelope at previous times, all the way to the epoch when the ejecta was fully ionized.
This calls for a change of strategy in SN spectroscopic modeling, and casts doubt
on results using approaches that ignore the history of the SN ejecta.
While the energy gain through hydrogen recombination in the outflow is a feature of
Type II SN ejecta only, the frozen-in ionization at large distances/velocities
stems from the order-of-magnitude larger recombination time-scale compared to expansion 
time-scale. This fact is not influenced by the energetics of recombination ---
it merely demonstrates that SN ejecta are so tenuous and so fast
that the ionization balance can be considerably affected by time dependence.
This phenomenon should be a ubiquitous feature of SN ejecta and, at the very least, deserves
considerably more attention that it has so far received.

Accounting for time dependence in CMFGEN opens new horizons. In particular, it is now
sensible to model Type II SN throughout the photospheric phase.
We will incorporate in CMFGEN the energy deposition due to unstable isotopes,
to cover the nebular phase, and investigate a wider range of SN explosions,
from core-collapse to thermonuclear runaways. We will also include relativistic effects, and
time dependence for the radiation transport, to improve the energy transport at depth.
We will finally assess the effects of time dependence on the correction factors used in 
the Expanding Photosphere Method \citep{DH2005b}.

%
%
%
%




\bibliographystyle{aipproc}   

\bibliography{ddt}

\IfFileExists{ddt.bbl}{}
 {\typeout{}
  \typeout{******************************************}
  \typeout{** Please run "bibtex ddt" to optain}
  \typeout{** the bibliography and then re-run LaTeX}
  \typeout{** twice to fix the references!}
  \typeout{******************************************}
  \typeout{}
 }

\end{document}